\documentclass[12pt,preprint]{aastex}

\usepackage{graphicx}
\usepackage{natbib}

\usepackage{amssymb}
\usepackage[figuresright]{rotating}

\newcommand{\gapprox}{$\stackrel {>}{_{\sim}}$}   

\slugcomment{To appear in Ap. J.}

\begin{document}

\title{On the mid-IR variability of candidate eruptive variables (EXors):\\
a comparison between {\it Spitzer} and {\it WISE} data}

\author{S. Antoniucci\altaffilmark{1},
T. Giannini\altaffilmark{1},
G. Li Causi\altaffilmark{1}, 
and
D. Lorenzetti\altaffilmark{1}
}
\altaffiltext{1}{INAF - Osservatorio Astronomico di Roma, via
Frascati 33, 00040 Monte Porzio, Italy, 
simone.antoniucci@oa-roma.inaf.it, teresa.giannini, gianluca.licausi, dario.lorenzetti}

\begin{abstract}
Aiming to statistically study the variability in the mid-IR of young stellar objects, we have compared the 3.6, 4.5, and 24 $\mu$m {\it Spitzer} fluxes of 1478 sources
belonging to the C2D (Cores to Disks) legacy program with the {\it WISE} fluxes at 3.4, 4.6, and 22 $\mu$m. From this comparison we have selected a robust sample of 34 variable sources. Their variations were classified per spectral Class (according to the widely accepted scheme of Class I/flat/II/III protostars), and per star forming region. On average, the number of variable sources decreases with increasing Class and is definitely higher in Perseus and Ophiuchus than in Chamaeleon and Lupus. According to the paradigm Class $\equiv$ Evolution, the photometric variability can be considered to be a feature more pronounced in less evolved protostars, and, as such, related to accretion processes. Moreover, our statistical findings agree with the current knowledge of the star formation activity in different regions. The 34 selected variables were further investigated for similarities with known young eruptive variables, namely the EXors. In particular we analyzed : (1) the shape of the spectral energy distribution (SED); (2) the IR excess over the stellar photosphere; (3) magnitude versus color variations; and (4) output parameters of model fitting. This first systematic search for EXors ends up with 11 {\it bona fide} candidates that can be considered as suitable targets for monitoring or future investigations.
\end{abstract}


\section{Introduction}

Flux variability is a typical feature of almost all Young Stellar Objects (YSOs); it is detected over a wide interval of wavelengths (more than 4 dex) from X-rays (e.g. Audard et al. 2005, 2010; Grosso et al. 2005, 2010) to mid- to far-infrared (IR) (e.g. Rebull et al. 2011, Faesi et al. 2012; Hillenbrand et al. 2013; Antoniucci  Giannini \& Lorenzetti, 2013c). Each spectral band, where variability is investigated, provides information on specific locations of the whole system (central star, 
circumstellar disk, external envelope) at different spatial scales ranging from 0.1 to 100 AU. Fluctuations typical of YSOs are modest (0.2-0.4 mag) but significant, from the optical to the mid-IR, and occur on daily to monthly time-scales. They are due to a wide variety of physical mechanisms, including variations in both the accretion rate and dust extinction. Different manifestations of these processes  are the rare EXor, FUor, or UXor events. Disk accretion phenomena are characterized by intermittent outbursts (usually detected in the optical and near-IR bands) due to the sudden increase of the mass accretion rate (by orders of magnitude, Hartmann \& Kenyon 1985). D'Angelo \& Spruit (2010) recently provided quantitative predictions for the episodic accretion on magnetized stars and
indicated that the cycle time of the bursts increases with a decreasing accretion rate. Such events are typical of many (if not all) YSOs. The outbursts of larger intensity (\gapprox~4 mag) are classified into two major classes: ({\it i}) EXor events (Herbig 1989) lasting one year or less, with a recurrence time of months to years, characterized by emission line spectra; and ({\it ii}) FUor events (Hartmann \& Kenyon 1985) of longer duration (\gapprox~tens of years)
with spectra dominated by absorption lines. Although we are mainly interested in accretion-driven variability, it is worth mentioning that there exists a different type of variability (extinction-driven), which presents observational features similar to those of many accreting objects. This type of photometric activity is defining the UX Ori stars characterized by periodic decreases of luminosity due to the transit of dust clumps (enhancements of the circumstellar extinction), which are observed in the optical bands and, with decreasing amplitude, in the IR ones. Observational results provide evidence that the UXor variability is a consequence of instability of the innermost layers of their gas/dust accretion disks (Shenavrin et al. 2012).\

Monitoring programs to investigate the YSOs' variability are typically carried out in the visual and near-IR bands (0.4-2.5 $\mu$m) that are the most easily accessible from the ground (see e.g. Lorenzetti et al. 2012, and references therein). So far, these surveys have left practically unexplored the range of wavelengths greater than 3 $\mu$m (a recent census of the available mid-IR studies is given by Faesi et al., 2012). A strong impulse has been given by {\it Spitzer} (Werner et al. 2004) and in particular by the program YSOVAR (Young Stellar Object VARiability), the first large-scale systematic survey of YSOs' mid-IR variability (Morales-Calder\'{o}n et al. 2011). In the Orion Nebula Cluster (ONC) they found a wide variety of variability types and a correlation between the evolutionary stage (Class I/II/III) and the amplitude of the fluctuations that is more pronounced for younger objects. The increasing interest toward the mid-IR domain (3-25 $\mu$m) is mainly due to the fact that the spectral behavior of YSOs at these wavelengths is strictly related to disk and envelope regions located at radial distances (from the central star) where disk fragmentation and planet formation occur. This latter circumstance is signaled by the presence of inner holes in the circumstellar disks, a scenario largely supported by both photometric (Sipos et al. 2009, Lorenzetti et al. 2012) and interferometric (Akeson et al. 2005, Eisner et al. 2009) studies. 

Recently, between 2010 and 2011, the space mission {\it WISE} (Wright et al. 2010) has photometrically covered the whole sky by using filters whose band-passes are very similar to those of {\it Spitzer}. This circumstance allows us to conduct an unbiased study of the mid-IR variability (on a 5 years time-scale) of all the YSOs defined by the {\it Spitzer} C2D legacy program (Evans et al. 2009) and belonging to the nearby star-forming regions. In particular we aim at searching for similarities (or differences) between the photometric behavior of known EXor variables and those of randomly variable YSOs. Because {\it Spitzer} and {\it WISE} data provide two sets of photometric measurements taken in two different epochs, they offer the chance to perform a statistically significant comparison between YSOs belonging to the same cloud. Recently, Scholz, Froebrich \& Wood (2013) presented a systematic survey of eruptive YSOs analogously based on the comparison {\it Spitzer} versus {\it WISE}. In their work however, they looked for the rarer FUor outbursts and therefore their methods of selection are different from ours (see Section 3.3 for further details). 

The present paper is organized as follows: the investigated sample is defined in Section 2, while in Sectoin 3 the criteria to identify variable objects are provided along with statistical considerations on the obtained objects. In Section 4 our results
are presented and discussed in terms of the existing accretion-driven and/or extinction-driven scenario. Finally, our concluding remarks are given in Section 5.

\section{The investigated samples}

\subsection{The YSOs Sample}

Our comparison ({\it Spitzer} versus {\it WISE}) is justified because both space-crafts have similar performances in terms of sensitivity, resolution, and saturation limit (see Table~\ref{properties:tab}). Of course {\it Spitzer} data are only available for a small fraction of the all-sky area covered by {\it WISE}, but, when available, the {\it Spitzer} 85 cm primary mirror (versus 40 cm for {\it WISE}) provides an image resolution twice as good as that achieved by {\it WISE} in similar bands. The limiting magnitudes of our sample are also given in Table~\ref{properties:tab}.\ 

In this paper we make use of the WISE All-Sky Source Catalog and of the catalogs provided by the C2D {\it Spitzer} legacy program (Evans et al. 2003) dealing with 
YSOs and candidates carefully selected and classified by means of their colors and spectral energy distributions (SEDs) 
(Evans et al. 2009). In particular, we used the final C2D data release (DR4, Fall 2006 and Fall 2007) and retrieved 
archival data for the C2D sub-samples named (YSO)CLOUDS, (YSO)OFF-CLOUDS, (YSO)CORES, and (YSO)STARS, which include all the 
sources from the full catalog that are classified as YSO candidates (http://peggysue.as.utexas.edu/SIRTF/). In practice, 
our starting sample coincides with that of Scholz, Froebrich \& Wood (2013), but the applied constraints and selection 
criteria for variables (see below) are different, so that we end up with a different selected sub-sample.

In total we start with 1478 sources whose vast majority belongs to nearby star forming regions (Oph, Lup, Cha, Per, Ser); few complementary objects are located in Taurus and in other nearby dark cores. Aiming at comparing {\it Spitzer} with {\it WISE}, we are concerned with {\it Spitzer} bands at 3.6 $\mu$m (IRAC1), 4.5 $\mu$m (IRAC2), and 24 $\mu$m (MIPS1), whose central wavelengths are very similar to those of {\it WISE} W1, W2, and W4 bands (at 3.4, 4.6, and 22 $\mu$m, respectively).\ Hence, we will compare fluxes in these three bands, which will be called in the following B1, B2, and B3, respectively. 

For each region, IRAC data of the C2D survey were taken at two different times (\textit{epochs}) typically separated by a few hours. The search for variability over the short time-scales between the two epochs has given negative (in Serpens by Harvey et al. 2007 ApJ 663, 1149) or inconclusive (in Chamaeleon I by Luhman et al. 2008 ApJ 675, 1375) results. Thus, in the present work we have considered the provided combined photometries between the two epochs.
In this context, it is remarkable that the variability item was never addressed in other papers dealing with the C2D survey. Instead, two other regions were surveyed by IRAC in two epochs separated by six months: Orion (Megeath et al. 2012) and Vela Molecular Ridge (Giannini et al.
2009). Although they use different scopes and selection criteria, both works investigate source variability, finding that many objects present significant fluctuations. This circumstance signals that variability is more likely associated to longer time-scales (months),
which are those we investigate here.

\begin{figure}
\includegraphics[angle=0,width=16cm]{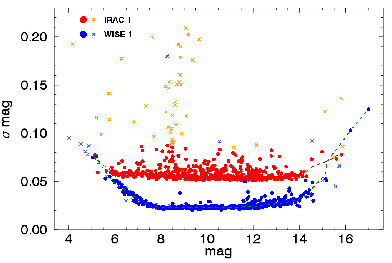}
   \caption{Distribution of the B1 photometric errors of the sources of our sample as a function of the B1 magnitude detected by {\it Spitzer} (red circles) and  {\it WISE} (blue circles). The cross symbols represent data points that we removed from our sample for being outliers with respect to the median trend (dashed lines). 
   \label{fig1:fig}}
\end{figure}

Data taken in different, although adjacent, bands (e.g. IRAC4 at 8.0 $\mu$m and {\it WISE} W3 at 12 $\mu$m) have not been considered because of the large difference between their band-passes. For each considered {\it Spitzer} band, we selected the sources having a S/N$>$5 (namely those with a photometry quality flag equal to or better than {\it C}, according to the C2D flags). These sources were further inspected by plotting their error distribution as a function of the magnitude (Figure~\ref{fig1:fig}). Notably, we found that some sources in B1 show large outlying errors with respect to the global trend (clustered at about 8 mag in Figure~\ref{fig1:fig}). The same is true for B2 (not shown). 
A tentative explanation is that the results are relative to combined High Dynamic Range (HDR) IRAC observations in which stars around 8 mag are saturated in long exposure images, whereas they are fairly faint in short exposures and consequently show larger photometric errors (which are those eventually associated with these objects in the catalog). To be conservative, we decided to remove these outliers from our sample because they might significantly alter our results.

As a further step, we searched for the {\it WISE} counterparts of the {\it Spitzer} sources within a radius of 1 arcsec. After this operation, 1286 sources in common between the two catalogs (Table~\ref{sample:tab}) remained. The magnitude limit of our sample is imposed by {\it WISE} and corresponds to about 16.9, 14.0, and 9.4 mag at 3.4, 4.6, and 22 $\mu$m, respectively.  Among these, we eventually kept the ones simultaneously satisfying some additional constraints, based on the available {\it WISE} flags (Cutri et al. 2012): {\it (i)} S/N ratio$>$5 in W1, W2, and W4, to have sets of comparable quality; {\it (ii)} total percentage of saturated pixels less than 20\%; {\it (iii)} contamination flag = 0, indicating the presence of no spikes or ghosts; {\it (iv)} extendedness flag = 0, to select only point-like sources. 

Additionally, because we made use of \textit{Spitzer} IRAC4 and MIPS1 measurements to obtain magnitude corrections for B3 (see Section 2.1.1), we imposed an extra constraint for B3 data, namely that the time interval between IRAC4 and MIPS1 observations of a given target does not exceed seven days. This was required to minimize possible biases arising from flux variations occurring between the observation dates of the two \textit{Spitzer} instruments. In the following, we consider \textit{Spitzer} measurements fulfilling this constraint as basically simultaneous.

In principle, Asymptotic Giant Branch (AGB) stars colors at these wavelengths could partially overlap with those of YSOs, but AGB presence as background objects of star-forming regions (which, moreover, do not belong to the Galactic Plane) is very unlikely. 

At the end of this process we had 578, 480, and 73 sources in B1, B2, and B3\footnote{These represent about one third of the initial sample for B1 and B2; the lower  percentage found for B3 is due to the extra constraint on time interval.}, respectively, distributed among the YSOs classes as reported in (Table~\ref{sample:tab}). This is the final sample that we will consider for our variability analysis.

\subsubsection{Color Corrections}

{\it Spitzer} and {\it WISE} photometry were obtained with similar, but not identical filters (3.6, 4.5, 24 $\mu$m and 3.4, 4.6, 22 $\mu$m, respectively), so that the differential magnitude between the two measurements must be corrected to take into account the different effective wavelengths of the corresponding filters (e.g. Lorenzetti et al. 2012).

The relationships for the general corrections (which have been applied to the {\it Spitzer} magnitudes source by source) are:

\begin{equation}
   m_{\it Sz}(3.4\mu m) = m(IRAC1) - 0.20933\cdot[m(IRAC2)-m(IRAC1)]
\end{equation}
\begin{equation}
   m_{\it Sz}(4.6\mu m) = m(IRAC2) + 0.11644\cdot[m(IRAC2)-m(IRAC1)]
\end{equation}
\begin{equation}
   m_{\it Sz}(22\mu m) = m(MIPS1) - 0.10040\cdot[m(MIPS1)-m(IRAC4)]
\end{equation}

\noindent
which can be obtained by linearly interpolating (extrapolating for B1) the {\it Spitzer} magnitudes between those bands for which a "good" photometry is available (i.e. satisfying the constraints listed in previous section), and then computing the {\it Spitzer} flux at the effective wavelengths of {\it WISE} bands.

Finally, we added to previous relationships (1)-(3) the following magnitude shifts: 0.07146, -0.01855, -0.26179, respectively. These were 
computed as the $3\sigma$-clipped weighted mean of the magnitude difference over the entire sample and represent the magnitude shifts to be applied to the
data distribution in order to realign them, assuming a zero mean magnitude variation between the two photometric sets.

The final average corrections for the sources of our sample are $\sim$ 0.2 mag for B1, 0.05 mag for B2, and 0.5 mag for B3.

\subsection{The Control Sample} 

In order to validate our results, we defined a control sample for which we applied the same approach adopted for the investigated sample. In the following section we compare and discuss any differences between the two samples. The control sample was constituted from normal stars, namely by the 28,167 entries of the {\it Spitzer} High Reliability (HIREL) STARS Catalog. Noticeably, these objects are located in the same five clouds investigated by the C2D survey. The number of sources that are kept in the various bands after applying the same selection procedures described in Section 2 are reported in Table 4.

\section{Identification of variable sources}

\subsection{Selection Criteria}

The variable objects within a given band were conservatively defined as those presenting an absolute magnitude variation larger than 5 times the magnitude error:
$\mid$m$_1$ - m$_2$ $\mid$ $\geq$ 5$\sigma$(m$_1$-m$_2$), where m$_1$, m$_2$
refer to {\it WISE} and {\it Spitzer} magnitudes, respectively. Such a criterion warrants selecting genuine variations well above the photometric errors. 
We also remind the reader that all the considered detections are individually above 5$\sigma$ (see Section 2) to minimize any contamination by unwanted effects such as spikes and artifacts. 

\subsection{Statistical Considerations}

Statistical results about the detection rate of rising or declining variables, number of variables per Class (I/flat/II/III) and per cloud are given in Table~\ref{stat:tab}. Firstly, we note that the ratio between the number of variables and the total number of sources detected in a given band indicates that searching for variability in the first two bands provides a success rate definitely larger than in B3. Indeed, the percentage of variable sources in a given band decreases with increasing wavelength (from 16\% in B1 to about 3\% in B3). This percentage strongly diminishes when considering the simultaneous and concordant (i.e. with the same sign) variations in multiple bands: indeed, we do not find any source concordantly variable in all the three bands, an effect likely due to our stringent requirement of a variation $>$ 5$\sigma$ in each band. 
Therefore, because accretion objects usually present fluctuations whose amplitude decreases with the wavelength (Lorenzetti et al. 2007), we decided to focus on sources concordantly variable in the first two bands (34 objects). 

Secondly, in case of random fluctuations, the number of brightening sources (N$_{increased}$) should be similar to that of fading ones (N$_{decreased}$) this is not true in B1, where fading events exceed the brightening ones; it is marginally true in B2, where 
N$_{increased}$ $\sim$ N$_{decreased}$. 
Based on the number of brightening and fading sources in Vela, Giannini et al. (2009) found that the declining time could typically be longer than the rising one, by a factor of roughly 1.5, as observed in EXors light-curves (Lorenzetti et al. 2006).
This is in agreement with what we observed in B1, while we do not find the same evidence in B2. However, B1 is the most sensitive band to EXor-like variations thanks to the shortest effective wavelength.
As for B3, the very low number of variable objects (2 out of 73) does not allow us to provide significant statistical considerations.

\begin{figure}
\includegraphics[angle=0,width=16cm]{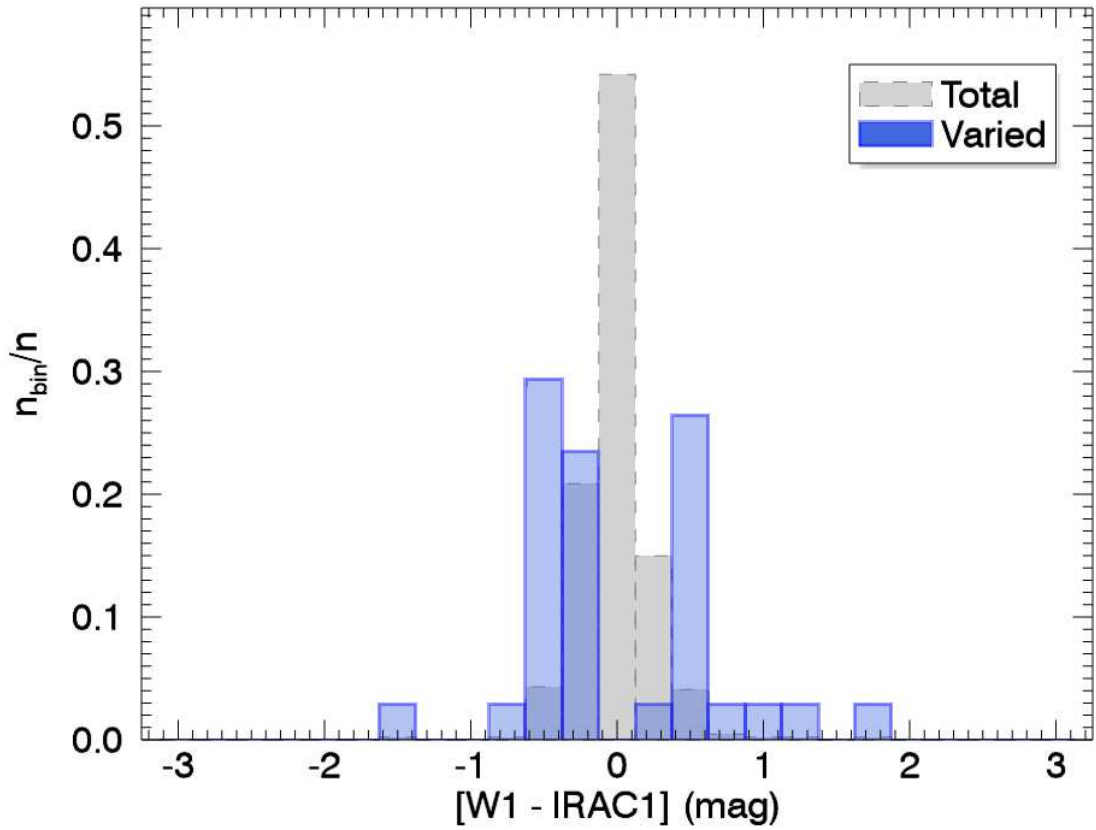}
   \caption{The normalized distribution of the B1 magnitude variation of the selected 34 sources concordantly varied in B1 and B2 (blue) is compared to the normalized distribution of the 441 sources having good B1 and B2 photometry (light gray).
   \label{histo:fig}}
\end{figure}

In Figure~\ref{histo:fig} the normalized distribution of variable sources concordantly variable in B1 and B2 (441 objects) is given as a function of the B1 magnitude variation. As naturally expected according to the events statistics, the relative number of variable objects in each bin decreases by increasing the amount of relative variation. 
Moreover, the distribution of variables does not present a sharp peak in correspondence of small variations ($|\Delta$mag$|<$ 0.1) as the total distribution does: this effect is caused by the selection criterion, which selects only those variations significantly larger than their own errors.\

The results obtained by applying the same analysis to the control sample are summarized in Table~\ref{stars:tab}. As expected, no significant variability affects this sample; indeed, this sample is dominated by Main Sequence (MS) stars that, as such, are less affected by IR variability, a phenomenon more typical of young protostars. However, the most striking and convincing difference between the YSOs and the control sample are evident when comparing color and magnitude variations (see Section 4.1).\

The data of Table~\ref{stat:tab} indicate that the percentage of variable YSOs decreases monotonically (in all the considered bands) when passing from Class I to Class III sources. The same result was also found by Morales-Calder\'{o}n et al. (2011), who extensively investigated a dozen
star-forming regions. In the context of the widely accepted paradigm, according to which different SED shapes identify different evolutionary stages, our statistics suggest that the relevance of the photometric variability diminishes while the object evolves. Hence, the mass accretion rate ($\dot{M}$) onto the central object should have a role in regulating the level of the observed variability: the relative importance of the latter progressively fades with $\dot{M}$ value, from 10$^{-6}$-10$^{-7}$ (ClassI/flat) to 10$^{-7}$-10$^{-10}$ M$_{\sun}$ yr$^{-1}$ (ClassII/III).
By examining the data of Table~\ref{stat:tab} relative to any individual star forming region, Perseus appears to be the most active (from the variability point of view), followed by Ophiuchus and Serpens, whereas Lupus and Chamaeleon are fairly less active. Indeed, this scenario agrees very well with the current estimates of the star formation rates of these regions (e.g. Enoch et al. 2009; Evans et al. 2009). 
These considerations suggest that the type of variability evidenced here is probably ascribable in large part to accretion processes.

\subsection{Final Sample of Variable YSOs}

Among the variable objects that are significantly variable in B1, 34 sources are also simultaneously and concordantly variable in both B1 and B2 (Table~\ref{stat:tab}). Because the known EXors are always observed to vary in more than one spectral band, this latter sub-sample is well suited to our purpose, even if it only represents about the 8\% of the total number of objects detected in both bands. A catalog of these 34 sources is given in Table~\ref{selected:tab}, where the identification of the closest SIMBAD counterpart (typically within 0.1-0.2 arcsec) is listed, as well. 

The percentage of variables we found (34/441) may appear fairly low when compared to previous works, for instance the YSOVAR project (Morales-Calder\'on et al. 2011) has found a percentage of variables around 70\%.
This is due to several reasons that cannot be exactly quantified or easily disentangled. 
The most important difference is that the YSOVAR variables refer only to a peculiar region (ONC) whose population is dominated by younger objects (with respect to other star-forming regions), which show on average a much more pronounced variability. In addition, the variables found by YSOVAR also include objects showing small magnitude variations ($|\Delta$mag$|<$ 0.2), which are the most frequent but tend to be disregarded with our stringent selection criteria (Section 2.1).
Additionally, we note that the variability index employed by YSOVAR is evaluated on 40 different epochs instead of only 2, and the sensitivity of such index increases as $n^{1/2}$, where $n$ is the number of the epochs. Finally, the method used in YSOVAR assumes that the photometric errors in the different bands are uncorrelated, which is not always verified. 


Our results can also be compared to conclusions by Sholz et al. (2013), who found no candidate eruptive variable starting from the same catalog of 1478 objects. These results were not inconsistent and only depend on the different selection criteria and aims of the two works.
Indeed, being interested only in strongly outbursting YSOs to look for FUOr-like objects, the main selection rule adopted by Scholz et al. is to have mag(W1-IRAC1) $< -1$ in both B1 and B2. Noticeably, none of the 34 objects in our final selected sample satisfies this request (the source \#192 shows $\Delta$mag $< -1$, but only in B1), so they were not selected in the Scholz et al. procedure. 
Interestingly, we also infer that all the 23 objects considered in the final list of Scholz et al. and then discarded (on the basis of conditions like bad association, saturation, or extendedness) are likely rejected by our selection criteria based on WISE flags (Section ~2.1).

All the 34 sources were accurately scrutinized to verify weather their variation was genuine (i.e. not due to fluctuations of the local sky or background between the two epochs), and to reject galaxies or close companions not resolved by {\it WISE}. To this scope we have checked the [{\it WISE}-{\it Spitzer}] magnitudes in B1 and B2 of all the IRAC sources within 1$\arcmin$ from the selected variable. Indeed, we verified that no other source exists (within 1$\arcmin$) that satisfies our variability criteria.

\section{Results and discussion}

\subsection{Colors versus Magnitude Variations}

\begin{figure}
\includegraphics[angle=0,width=16cm]{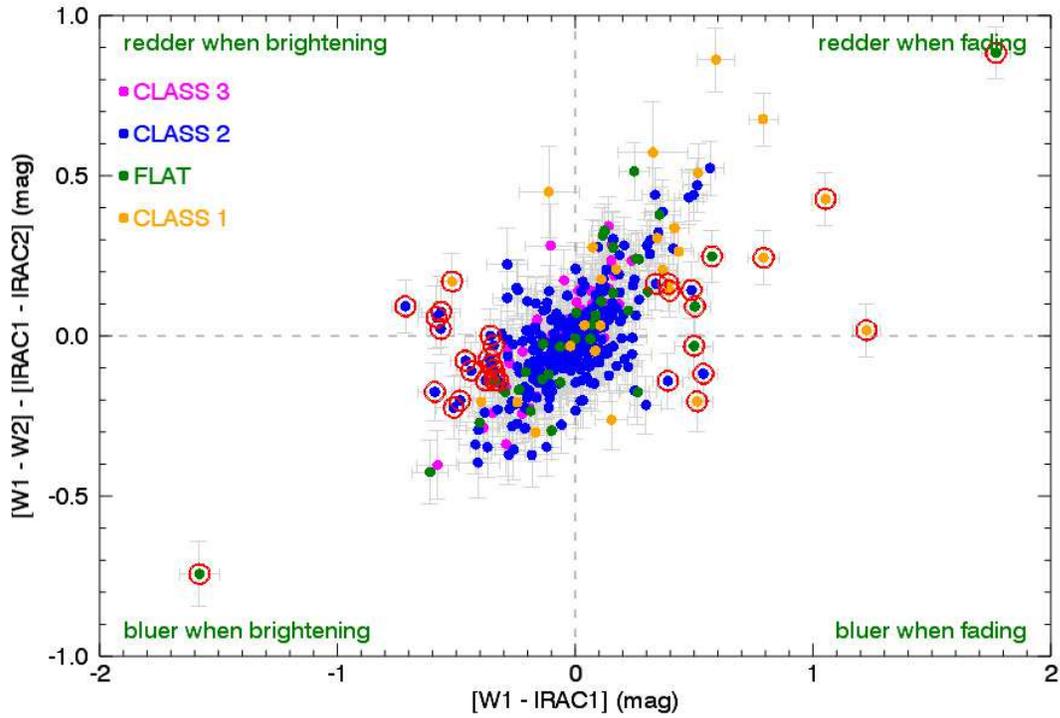}
   \caption{Color-magnitude variations ([W1--W2]--[IRAC1--IRAC2] and W1--IRAC1) of the 441 sources with good B1 and B2 photometry. Different colors indicate different Classes, as outlined in the legend, while red-circled data-points indicate the selected 34 sources concordantly varied in B1 and B2. 
   \label{histo_ysos:fig}}
\end{figure}

Figure~\ref{histo_ysos:fig} depicts the color variations [(B1-B2)$_{2nd~epoch}$ - (B1-B2)$_{1st~epoch}$] as a function of the magnitude variation (in B1) for all 441 selected sources of our sample observed both in B1 and B2 (see Table~\ref{selected:tab}). Different colors indicate different Classes, as outlined in the legend. The red-circled sources are those presenting $\Delta$mag variations above 5$\sigma$ and concordantly variable in B1 and B2 (see Section 3.3), namely the 34 sources presented in Section 3.3. 

\begin{figure}
\includegraphics[angle=0,width=16cm]{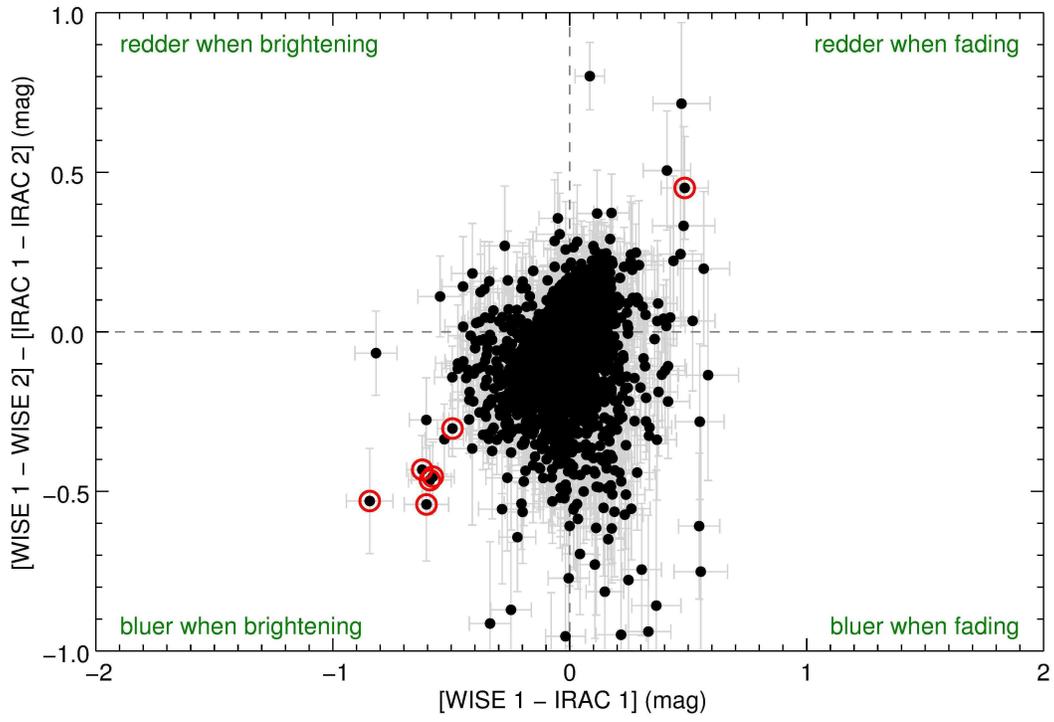}
   \caption{As Figure~\ref{histo_ysos:fig} for the control sample of HIREL STARS.
   \label{histo_stars:fig}}
\end{figure}

Noticeably, when searching for {\it bona fide} EXor (or FUor) candidates, we did not impose any additional selection criteria on the color variation. 
Indeed, although passing from quiescence to outburst all EXors move to bluer positions in a two-colors plot (Lorenzetti et al. 2012), in a few cases variations have been observed in just one color. For instance, this is the case of V2493 Cyg (alias HBC722) presented in Figure~5 of Lorenzetti et al. (2012): if we had imposed some selection criteria on the variation of the [H-K] color, we would have ignored this source that, conversely, has been considered to be one of the most interesting EXor currently under investigation (Miller et al. 2011; K\'{o}sp\'{a}l et al. 2011; Green et al. 2013; Antoniucci et al. 2013a and references therein).
The four quadrants of Figure~\ref{histo_ysos:fig} are differently labeled: the 1st and 3rd quadrants, identify the {\it loci} corresponding to sources that become bluer when brightening, whereas the 2nd and 4th quadrants refer to those that become redder when brightening. Obviously the great majority of sources do not present significant variations either in magnitude or in color and thus, they tend to cluster around the origin of the axes. However, an evolutionary trend according to which larger variations are associated to the younger sources, is recognizable. Indeed, the median ($\Delta$mag,$\Delta$col) are (0.37,0.21), (0.21,0.14), (0.10,0.07), and (0.10,0.08) mag for the 27 Class I, 35 flat-spectrum, 312 Class II, and 67 Class III sources, respectively (see also Table 2). The same plot relative to our control sample is depicted in Figure~\ref{histo_stars:fig},
where the distribution of the HREL STARS does not present any peculiarity, at variance with the YSOs distribution (Figure~\ref{histo_ysos:fig}), which is definitely 
oriented along the 1st and 3rd quadrants, as expected for young sources whose variability is likely driven by accretion (or extinction) fluctuations.\ 

In Lorenzetti et al. (2012) we noticed a very similar distribution by comparing a sample of T Tauri stars observed in two different epochs: namely Cohen \& Kuhi (1979) and 2MASS catalogs. Although that comparison concerned the [H-K] color variation versus $\Delta$H magnitude variation, instead of the present [B1-B2] versus B1 variation, the observed pattern and also the spread (both in magnitude and color) are substantially similar, thus suggesting that we are evidencing an actually intrinsic property of young objects.

\subsection{SEDs and Similarities with EXors}

In Section\,3.3 we have identified 34 sources with a significant and concordant variation in both B1 and B2. These sources represent a sample suitable for a search of candidate EXors.
We searched for similarities with known EXors, looking in particular for objects that satisfy the following requests: (1) to show a remarkable IR excess in the SED shape, and, (2) to become redder when fading (or bluer when brightening).

\begin{figure}
\includegraphics[angle=0,width=17cm]{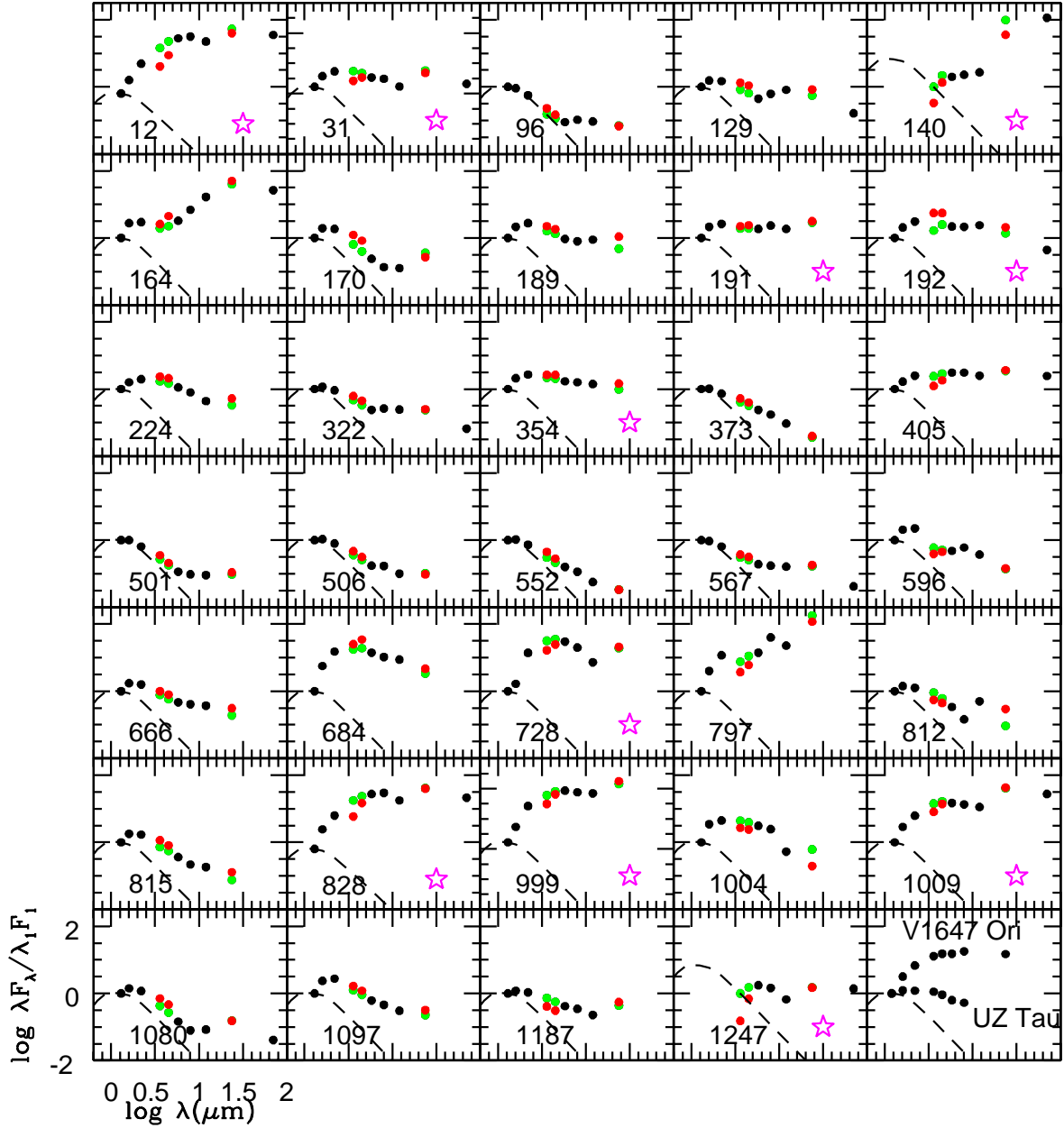}
   \caption{SED's of the 34 variables in the range 1.25-70 $\mu$m. Black dots refer to 2MASS (JHK) and 5.8, 8, 12, and 70 $\mu$m photometry. Green and red dots indicate {\it Spitzer} (IRAC1,2 - MIPS1) and {\it WISE} (W1, W2, and W3) data, respectively. The dashed line represents a median photosphere of stars in the spectral range K5-M5 and normalized to the J band flux of each source, or at 3.4 $\mu$m if the source is undetected in J. The 11 sources eventually selected as EXor candidates are marked with a magenta symbol. For comparison's sake, the SED's of two EXor sources, V1647 Ori (embedded) and UZ Tau (classical) are depicted in the last panel.   
\label{sed:fig}}
\end{figure}

The near- and mid-IR SEDs are shown in Figure~\ref{sed:fig}. These have been constructed  by using {\it Spitzer}/{\it WISE} data along with their 2MASS counterparts that match our sources within 2 arcsec or less. {\it Spitzer} ({\it WISE}) photometry in band 1, 2, and 3 corresponds to green (red) dots, whereas black dots refer to both 2MASS (JHK) and {\it Spitzer}/{\it WISE} (5.8, 8.0, 12, and 70 $\mu$m) data. Unfortunately, a single epoch SED cannot be obtained because of the temporal intervals lapsed between the three surveys. Here, we intend just to emphasize possible similarities between the selected variables and the EXor systems. To this aim, the typical SEDs of the two EXors V1647 Ori and UZ Tau (the former less embedded than the latter) are also given in the last panel of Figure~\ref{sed:fig}. To give a quantitative evaluation of the presented SEDs, we introduce, as an empirical indicator, the parameter $\varepsilon$, that is the ratio of the total luminosity Lum$_{IR}$ (computed between 1 $\mu$m and band 3) to the luminosity of a median K5-M5 photosphere (computed in the same wavelength range). In this first-order approximation the J-band flux is assumed to be entirely photospheric, as implied by the normalization used to show the SEDs in Figure~\ref{sed:fig}. The fact that the 34 sources are all low luminosity objects (see the Lum$_{IR}$ values in the 14th column of Table~\ref{selected:tab}) is a selection effect due to the original definition of the C2D sample that is constituted only by low-mass objects. The values of $\varepsilon$ thus obtained are given in the 15th column of Table~\ref{selected:tab}. We found (Antoniucci et al. 2013c) that the EXors are characterized by $\varepsilon$ values that roughly span one order of magnitude, between 10 (typical of classical EXors) and 10$^2$ (typical of more embedded EXors). The differences between the two sub-classes (classical and embedded) are qualitatively accounted for in Lorenzetti et al. (2012). Following this prescription, we consider as sources fulfilling the request (1) 16 objects that have $\varepsilon$~$>$~10. 
The request (2) is satisfied by sources coded as "YES" in the 16th column of Table~\ref{selected:tab}.

By combining (1) and (2) we end up with 11 sources that can be considered as good EXor candidates (bold-faced in Table~\ref{selected:tab}). 
Of course, the photometric variations detected in these objects might also be related in principle to accretion variability that does not necessarily imply the presence of EXor events (as observed in many active T Tauri Stars) and to extinction effects. 
In the latter case, however, the observed color and magnitude variations are compatible with extinction effects only in 2 candidate sources, namely \#191 and \#1009. In these objects the variations occur along the extinction vector in a 
B1-B2 versus B1
color-magnitude diagram and are consistent with an $A_\mathrm{V}$ change of about 5 mag.
Further monitoring and investigations of these objects at shorter wavelengths are definitely required to ascertain their EXor-like behavior.

As a final note, we remark that these 11 sources represent about the 2.5\% of the initial sample of YSOs (441 objects, i.e. those detected simultaneously in B1 and B2 in both the {\it Spitzer} and {\it WISE} epochs, see Table~\ref{sample:tab}). This percentage should be equal to the probability to detect a source once in burst and once in quiescence.
If the interval between two bursts is $\Delta$T and the duration of the burst is $\delta$T (with $\delta$T $\ll$ $\Delta$T) this probability is P=($\delta$T/$\Delta$T)*(1-$\delta$T/$\Delta$T). Taking $\delta$T in the range 0.5-1 yr (as typically observed in the light curves of known EXors), we get P=2.5\% if $\Delta$T ranges between 12 and 39 yrs, respectively, in good agreement with the typical recurrence of EXors bursts. 
The near- to mid-IR photometry of the 11 EXor candidates is given in Table~\ref{phot:tab}.

\subsection{Stellar Parameters}

\begin{figure}
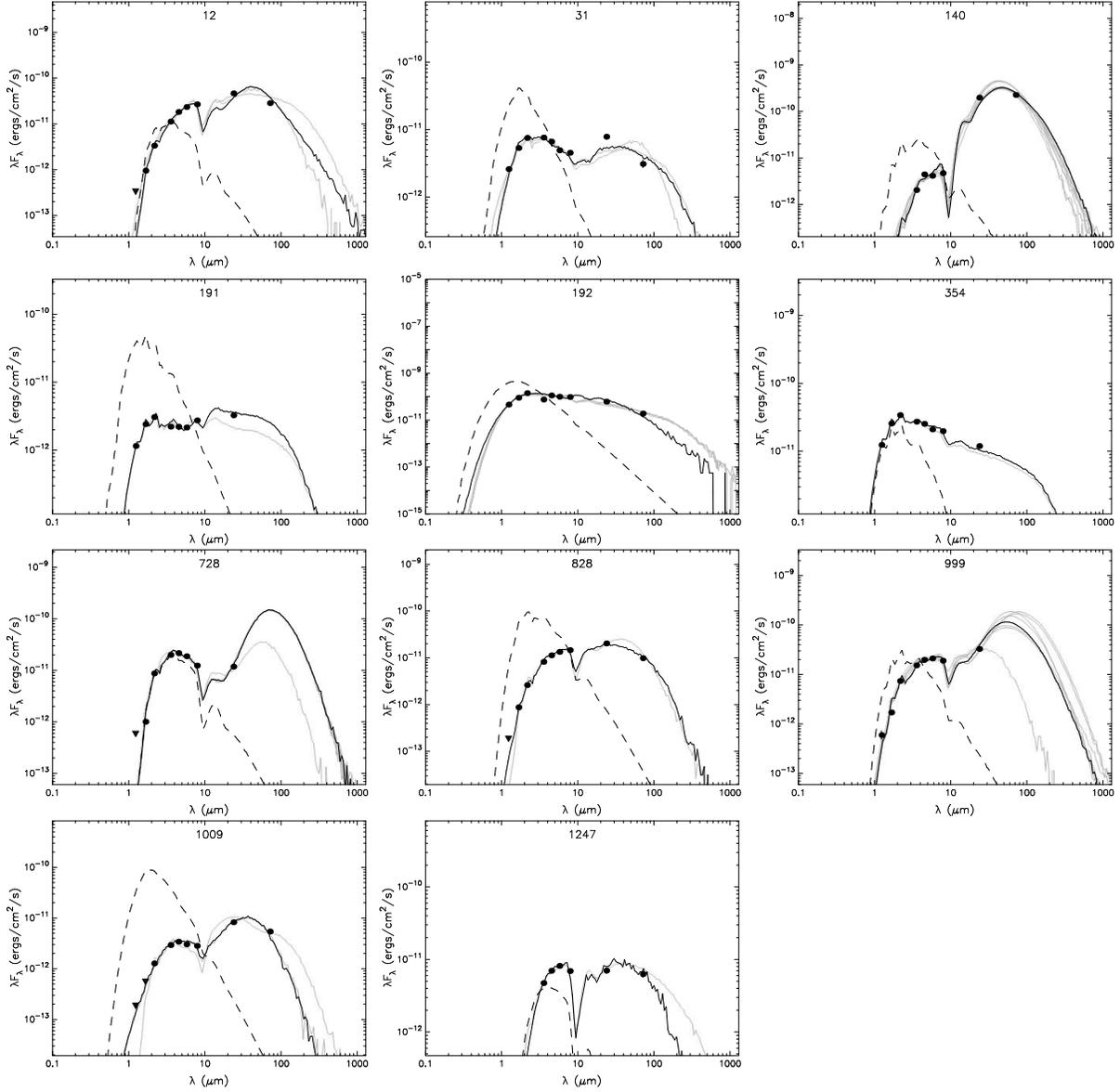


\includegraphics[angle=0,width=5.2cm]{12.eps}
\includegraphics[angle=0,width=5.2cm]{31.eps}
\includegraphics[angle=0,width=5.2cm]{140.eps}\\
\includegraphics[angle=0,width=5.2cm]{191.eps}
\includegraphics[angle=0,width=5.2cm]{192.eps}
\includegraphics[angle=0,width=5.2cm]{354.eps}\\
\includegraphics[angle=0,width=5.2cm]{728.eps}
\includegraphics[angle=0,width=5.2cm]{828.eps}
\includegraphics[angle=0,width=5.2cm]{999.eps}\\
\includegraphics[angle=0,width=5.2cm]{1009.eps}
\includegraphics[angle=0,width=5.2cm]{1247.eps}

	\caption{Best-fits of the 11 selected sources. The filled circles show the input fluxes, while triangles indicates 1$\sigma$  upper limits. The darkest line shows the best fit, while the grey lines show all the subsequent good fits with a $\chi^2$ up to 20\% higher than the minimum value. The dashed line shows the stellar photosphere corresponding to the central source of the best fitting model, as it would look in the absence of circumstellar dust (but including interstellar extinction). \label{fits:fig}}
\end{figure}

Once the SEDs of our variables have been constructed, we used the radiative transfer model developed by Whitney et al. (2003a, 2003b) to obtain insights on the physical properties of these objects. The community has been greatly facilitated in using this model  by Robitaille et al. (2006, 2007b), who provided a grid of 200,000 computed models and an efficient online fitting procedure. The results of the best-fits for the 11 selected sources are depicted in Figure~\ref{fits:fig}) and appear satisfactory. In the best-fit procedure we set as constraints the interstellar extinction between 0 and 30 mag and the appropriate distance $\pm$ 20 pc of the individual region (given in Table~\ref{selected:tab}). We used 2MASS, IRAC, and MIPS fluxes and different-sized apertures with a radius corresponding to 3$\sigma$ of the instrument beam and verified that even adopting larger apertures the same models are selected. The main output parameters are provided (for the 11 selected EXors) in Table~\ref{epsilon:tab}, where we give a pair of values: the former is relative to the best-fit model and the latter to the model with a $\chi^2$ 20\% higher than the minimum value. These parameters are the disk inclination (between 4$^{\circ}$ and 90$^{\circ}$ with respect to the sky-plane, (column 2), the interstellar (IS) and circumstellar (CS) extinction (columns 3, 4), the stellar mass and temperature (columns 5, 6), the outer envelope mass (column 7), and the total luminosity (column 8).\
 
The global picture emerging from the SED modeling of these 11 candidates is that they are low mass and relatively young protostars (Class I or flat SED). The stellar temperatures (2900$<T<$5200 K, although for sources \#192 and \#1247 higher temperatures appear possible) and masses (0.2-$\sim$2 M$_{\sun}$) are those typical of embedded (A$_V$ $\sim$ 10-50 mag) T Tauri stars of spectral type between K0 and M5. The reservoir of material of the surrounding envelope (M$_{env}$ ranging between 10$^{-8}$ and 10$^{-2}$ M$_{\sun}$ is not enough to significantly enhance the final stellar mass.

\section{Final remarks and perspectives}

We matched and compared the mid-IR {\it Spitzer} photometry of a sample of 1478 YSOs (from the C2D legacy program) with {\it WISE} data obtained about five years later with very similar spectral bands. Color corrections needed to convert {\it Spitzer} fluxes into {\it WISE} band-passes were applied and are provided. With a final scope of selecting potential EXor sources, we have considered the following steps:
\begin{itemize}
\item[-] Compelling criteria have been applied to both the {\it Spitzer} and {\it WISE} database to consider only unsaturated and point-like sources detected at S/N$>$5, which are also not contaminated by possible spikes and ghosts.
\item[-] YSOs present magnitude versus color variations characterized by a systematic trend of becoming redder when fading, whereas our control sample (constituted by normal stars) does not show any sign of peculiar pattern.
\item[-] Our YSOs have been grouped both per Class (following the protostars classification scheme ClassI/\-flat/\-II/\-III) and per star-forming region; on average,
the number of variable sources decreases by increasing the Class and is definitely higher in Perseus and Ophiucus than in Chamaeleon and Lupus. 
Therefore, the photometric variability is a feature more pronounced in less evolved protostars, and our statistical findings well agree with the 
current estimates of the star formation activity in those clouds.
\item[-] A more stringent constraint on the variability was imposed, first defining as true variables the sources presenting magnitude variations larger than five times the propagated magnitude error,
then selecting sources with concordant variations at 3.4 and 4.5 $\mu$m. We end up with 34 objects, 11 of which present similarities with known EXors, although we cannot rule out effects due to normal accretion variability (as observed in some active T Tauri Stars) or to variable extinction.

The number of candidates represents a large percentage (about 50\%) of the EXors known so far, a result showing the great potential of investigating large photometric databases.
\item[-] Model fitting of the SEDs of the 11 candidates provides output parameters (such as A$_V$, M$_{star}$, T$_{star}$, M$_{envelope}$, and total luminosity) well compatible with low mass, fairly evolved protostars, where disk accretion is expected to have a prominent role.  
\end{itemize}
Six out of the eleven candidates have a K-band magnitude $<$13, so they are accessible to the instrumentation we are using for our EXor monitoring program dubbed EXORCISM (EXOR optiCal Infrared Systematic Monitoring, Antoniucci et al. 2013b). These candidate sources are being included in our list of monitored targets, with the aim to trace their photometric variations and possibly confirm their EXor nature. 

\section{Acknowledgements}
We are grateful to the anonymous referee for suggestions and comments that allowed us to improve the quality of the article.

This work is based (in part) on observations made with the Spitzer Space Telescope, which is operated by the Jet Propulsion Laboratory, California Institute of Technology under a contract with NASA. It also makes use of data products from the Wide-field Infrared Survey Explorer, which is a joint project of the University of California, Los Angeles, and the Jet Propulsion Laboratory/California Institute of Technology, funded by the National Aeronautics and Space Administration. Complementary data comes from the Two Micron All Sky Survey, a joint project of the University of Massachusetts and the Infrared Processing and Analysis Center/California Institute of Technology, funded by the National Aeronautics and Space Administration and the National Science Foundation. We also acknowledge use of the SIMBAD database.

{}

\newpage

\begin{deluxetable}{lccc}
\tabletypesize{\normalsize} \tablecaption{{\it Spitzer} and {\it WISE} instrumental properties. \label{properties:tab}}
\tablewidth{0pt}
\tablehead{Band      &  mag limit     & resolution   & saturation  \\
                     &   (mag)        & (arcsec/pxl) &  (mJy)      }
\startdata
IRAC1 (3.6 $\mu$m)   &  16.9          &    1.2       &    439      \\
IRAC2 (4.5 $\mu$m)   &  15.9          &    1.2       &    450      \\
MIPS (24 $\mu$m)     &  9.8           &    2.6       &   1400      \\
WISE W1 (3.4 $\mu$m) &  16.9          &    2.7       &    178      \\
WISE W2 (4.6 $\mu$m) &  14.0          &    2.7       &    359      \\
WISE W4 (22 $\mu$m)  &  9.4           &    5.5       &  12080      \\
\hline
\enddata

\tablecomments{~~The listed values are taken from both the {\it WISE} Explanatory Supplement (Cutri et al. 2012)
and Robitaille el al. 2007a.}
\end{deluxetable}

\begin{deluxetable}{lc}
\tabletypesize{\normalsize} \tablecaption{The YSO sample. Number of objects with ``good" (see text) photometry in the indicated band(s) is given. \label{sample:tab}}
\tablewidth{0pt}
\tablehead{  & Number of objects }
\startdata
Initial {\it Spitzer} YSO sources ~~~~~~~~~~~~~~~~~~~~~~  &  1478     \\
B1                               &  578      \\
B2                               &  480      \\
B3                               &  73       \\
{\bf B1+B2}                      & {\bf 441} \\
B1+B2+B3                         &  34       \\
\hline
B1+B2 \& Class I $^a$            &  27       \\
B1+B2 \& Flat  $^a$              &  35       \\
B1+B2 \& Class II $^a$           &  312      \\
B1+B2 \& Class III $^a$          &  67       \\
\hline
\enddata

\tablecomments{The sub-sample of 441 objects detected in B1+B2 is investigated to search for EXor candidates (see text).}
\tablenotetext{a}{Once defined the spectral slope as $\alpha = dLog(\lambda F_{\lambda}/dLog\lambda$, Class I objects have $\alpha\geq$0.3, flat-spectrum objects have -0.3$\leq\alpha<$0.3, Class II objects-1.6$\leq\alpha<$-0.3, and
Class III objects $\alpha<$-1.6.}

\end{deluxetable}

\begin{deluxetable}{lccc|c}
\tabletypesize{\normalsize} \tablecaption{YSOs varied by more than 5$\sigma$. \label{stat:tab}}
\tablewidth{0pt}
\tablehead{                  & B1            &    B2       &  B3          & {\bf B1+B2}$^a$   }
\startdata
\multicolumn{5}{c}{Overall} \\
\hline   
N$_{varied}$             & 92 - 15.9\%   & 42 - 8.8\%  &  2 - 2.7\%   & {\bf 34 - 7.7\%}   \\
N$_{increased}$          & 36 - 6.2\%    & 23 - 4.8\%  &  2 - 2.7\%   & {\bf 20 - 4.5\%}   \\
N$_{decreased}$          & 56 - 9.7\%    & 19 - 4.0\%  &  0 - 0.0\%   & {\bf 14 - 3.2\%}   \\
\hline                                                                            
\multicolumn{5}{c}{By Class} \\
\hline   
Class I                  & 19 - 48.7\%   &  7 - 25.9\% &  0 - 0.0\%   & {\bf 8  - 29.6\%}   \\
Flat                     & 17 - 32.7\%   &  6 - 16.7\% &  0 - 0.0\%   & {\bf 6  - 17.1\%}   \\
Class II                 & 54 - 13.3\%   & 26 - 7.7\%  &  2 - 4.1\%   & {\bf 20 -  6.4\%}  \\
Class III                &  2 -  2.4\%   &  3 - 3.8\%  &  0 - 0.0\%   & {\bf 0  -  0.0\%}   \\
\hline                                                                             
\multicolumn{5}{c}{By Region} \\
\hline   
Cha                      &  0 - 0.0\%    &  0 - 0.0\%  &  0 - 0.0\%   & {\bf 0  -  0.0\%}   \\
Per                      & 28 - 30.4\%   & 16 - 38.1\% &  0 - 0.0\%   & {\bf 12 - 35.3\%}  \\
Ser                      & 15 - 16.3\%   &  7 - 16.7\% &  2 - 100.0\% & {\bf 6  - 17.6\%}   \\
Lup                      &  4 -  4.3\%   &  3 - 7.1\%  &  0 - 0.0\%   & {\bf 2  -  5.9\%}   \\
Oph                      & 25 - 27.2\%   & 12 - 28.6\% &  0 - 0.0\%   & {\bf 7  - 20.6\%}  \\
Other                    & 20 - 21.8\%   &  4 - 9.5\%  &  0 - 0.0\%   & {\bf 7  - 20.6\%}  \\
\hline
\enddata
\tablecomments{The 34 variables in B1+B2 (out of 441 objects, see Table\,2) are those further searched for EXor candidates (see text).
The pair of values (N - \%) indicate, for each band, the number of objects and the percentage with respect to: the number of sources considered (see Table 2) (Overall section),
the number of sources of each Class (Class section), and the number of varied sources (N$_{varied}$) (Region section).}
\tablenotetext{a}{Values of B1+B2 column refer to objects that vary concordantly in the the two bands.}
\end{deluxetable}

\begin{deluxetable}{lcccc}
\tabletypesize{\normalsize} \tablecaption{HIREL STARS varied by more than 5$\sigma$ selected from an initial sample composed of 28167 objects. \label{stars:tab}}
\tablewidth{0pt}
\tablehead{          &  B1  &  B2  &  B3   &  B1+B2    }
\startdata
N$_{tot}$            & 9113    &  3662    &   308     &   2936            \\
N$_{varied}$         &  222    &    37    &    16     &     15            \\
N$_{increased}$      &  110    &    27    &    14     &     14            \\
N$_{decreased}$      &  112    &    10    &     2     &      1            \\
%
%

\enddata


\end{deluxetable}

\begin{deluxetable}{ccccccccccccc|ccc} \tablecaption{Selected variables. \label{selected:tab}} 
\tabletypesize{\tiny} 

\tablewidth{0pt}
\tablehead{
\# & Our & \multicolumn{3}{c}{$\alpha$(2000.0)$^a$}&\multicolumn{3}{c}{$\delta$(2000.0)$^a$}  & sep.$^b$        & Class  & Region$^c$ & Dist.$^d$ &  Ident.$^e$ & L$^f_{IR}$    & $\epsilon^g$ & redder \\
  & Cat. &  $^{h}$ & $^{m}$ & $^{s}$               &${\degr}$ & ${\arcmin}$ & ${\arcsec}$     &(${\arcsec}$)    &        &        & (pc)  &             & (L$_{\sun}$ &            & fading   }
\startdata
  1  & {\bf 12}   & 03  & 28  & 00.40 & +30  & 08  & 01.19 &  0.15  & I    & PER & 250 &  LDN1455 IRS 3         &  0.296  & 281   &  Y    \\
  2  & {\bf 31}   & 03  & 28  & 50.63 & +30  & 42  & 44.59 &  0.19  & FLAT & PER & 250 &  (1) 155               &  0.073  & 12.0  &  Y   \\
  3  & 96   & 03  & 29  & 28.90 & +30  & 58  & 41.81 &  0.22  & II   & PER & 250 &  (1) 248               &  0.079  & 1.7   &  Y    \\
  4  & 129  & 03  & 32  & 41.70 & +31  & 10  & 45.78 &  0.48  & II   & PER & 250 &  2M J03324171+3110461  &  0.285  & 5.3   &  N    \\  
  5  & {\bf 140}  & 03  & 33  & 27.31 & +31  & 07  & 10.18 &  0.05  & I    & PER & 250 &  (2) SMM J033345+31071 &  0.785  & 156   &  Y   \\ 
  6  & 164  & 03  & 42  & 02.19 & +31  & 48  & 02.03 &  0.07  & I    & PER & 250 &  2M J03420217+3148019  &  0.096  & 52.7  &  N   \\
  7  & 170  & 03  & 42  & 27.14 & +31  & 44  & 32.84 &  0.27  & II   & PER & 250 &  (1) 324               &  0.137  & 4.1   &  N    \\ 
  8  & 189  & 03  & 43  & 28.44 & +32  & 05  & 05.96 &  0.18  & II   & PER & 250 &  2M J03432845+3205058  &  0.037  & 7.9   &  Y    \\	
  9  & {\bf 191}  & 03  & 43  & 36.03 & +31  & 50  & 08.96 &  0.05  & FLAT & PER & 250 &  2M J03433602+3150089  &  0.034  & 12.3  &  Y   \\	
 10  & {\bf 192}  & 03  & 43  & 44.49 & +31  & 43  & 09.37 &  0.17  & FLAT & PER & 250 &  2M J03434449+3143092  &  1.341  & 12.7  &  Y   \\ 
 11  & 224  & 03  & 44  & 18.58 & +32  & 12  & 53.09 &  0.03  & II   & PER & 250 &  V902 Per              &  0.144  & 6.5   &  Y    \\ 
 12  & 322  & 03  & 45  & 25.14 & +32  & 09  & 30.34 &  0.16  & II   & PER & 250 &  2M J03452514+3209301  &  0.330  & 2.9   &  Y    \\  
 13  & {\bf 354}  & 04  & 41  & 08.27 & +25  & 56  & 07.26 &  0.22  & II   & ITG33 & 150 &  ITG 33A               &  0.058  & 10.5  &  Y   \\  
 14  & 373  & 05  & 44  & 25.97 & +09  & 07  & 00.70 &  0.06  & II   & B35 & 400 &  (3) 249               &  0.815  & 2.3   &  Y    \\ 
 15  & 405  & 09  & 22  & 14.49 & -45  & 46  & 31.42 &  ...   & FLAT & DC2694+30$^h$ & 300 & \nodata      &  0.103  & 15.0  &  N   \\ 
 16  & 501  & 15  & 56  & 02.09 & -36  & 55  & 28.49 &  0.21  & II   & RXJ1556.1-3655 & 163 &  2M J15560210-3655282  &  0.249  & 2.0   &  Y    \\ 
 17  & 506  & 15  & 59  & 16.46 & -41  & 57  & 10.48 &  0.24  & II   & Sz129 & 150 &  2M J15591647-4157102  &  0.373  & 2.3   &  Y    \\  
 18  & 552  & 16  & 08  & 36.17 & -39  & 23  & 02.63 &  0.16  & II   & RXJ1608.6-3922 & 176 &  V1094 Sco             &  0.460  & 2.1   &  Y    \\	 
 19  & 567  & 16  & 09  & 01.84 & -39  & 05  & 12.57 &  0.18  & II   &LUPIII&200 &  V908 Sco              &  0.435  & 2.2   &  Y    \\ 
 20  & 596  & 16  & 12  & 11.21 & -38  & 32  & 19.69 &  0.10  & II   &LUPIII&200 &  2M J16121120-3832197  &  0.242  & 5.2   &  Y    \\	   
 21  & 666  & 16  & 25  & 36.73 & -24  & 15  & 42.65 &  0.27  & II   & OPH & 125 &  2M J16253673-2415424  &  0.301  & 4.4   &  N    \\  
 22  & 684  & 16  & 26  & 18.86 & -24  & 28  & 19.82 &  0.15  & II   & OPH & 125 &  2M J16261886-2428196  &  0.637  & 64.2  &  N   \\ 
 23  & {\bf 728}  & 16  & 27  & 03.00 & -24  & 26  & 14.86 &  0.19  & FLAT & OPH & 125 &  2M J16270300-2426146  &  0.018  & 99.9  &  Y   \\  
 24  & 797  & 16  & 27  & 48.24 & -24  & 42  & 25.74 &  0.17  & I    & OPH & 125 &  2M J16274825-2442256  &  0.013  & 171   &  N   \\ 
 25  & 812  & 16  & 28  & 13.70 & -24  & 31  & 39.17 &  0.11  & II   & OPH & 125 &  2M J16281370-2431391  &  0.002  & 3.9   &  Y    \\  
 26  & 815  & 16  & 28  & 16.73 & -24  & 05  & 14.55 &  0.28  & II   & OPH & 125 &  2M J16281673-2405142  &  0.159  & 4.1   &  Y    \\  
 27  & {\bf 828}  & 16  & 28  & 57.88 & -24  & 40  & 55.11 &  0.41  & I    & OPH & 125 &  (1) Oph emb 18        &  0.021  & 254   &  Y   \\
 28  & {\bf 999}  & 18  & 28  & 44.03 & +00  & 53  & 37.70 &  0.22  & I    & SER & 260 &  2M J18284402+0053377  &  0.169  & 148   &  Y   \\  
 29  & 1004 & 18  & 28  & 46.15 & +00  & 03  & 01.56 &  0.28  & II   & SER & 260 &  2M J18284613+0003015  &  0.026  & 15.0  &  N   \\	
 30  & {\bf 1009} & 18  & 28  & 51.23 & +00  & 19  & 27.17 &  0.03  & I    & SER & 260 &  2M J18285123+0019271  &  0.037  & 75.8  &  Y   \\  
 31  & 1080 & 18  & 29  & 35.63 & +00  & 35  & 03.70 &  0.18  & II   & SER & 260 &  2M J18293563+0035035  &  0.211  & 2.8   &  N    \\  
 32  & 1097 & 18  & 29  & 45.04 & +00  & 35  & 26.60 &  0.23  & II   & SER & 260 &  2M J18294502+0035264  &  0.069  & 6.3   &  Y    \\	
 33  & 1187 & 18  & 30  & 23.44 & +01  & 05  & 04.58 &  0.05  & II   & SER & 260 &  2M J18302343+0105045  &  0.078  & 3.5   &  N    \\ 
 34  & {\bf 1247} & 22  & 29  & 33.41 & +75  & 13  & 16.21 &  0.48  & FLAT &L1251& 300 &  (5) 042               &  0.055  & 14.2  &  Y   \\ 
\enddata
\tablecomments{References to the Table: (1) Evans et al 2009; (2) Jorgensen et al. 2007; (3) Dolan \& Mathieu 1999; (4) Lee \& Myers 1999; (5) Dunham et al. 2008; (6) Makarov et al. 2007; (7) Kun \& Prusti 1993; (8) Caselli et al. 2002.}
\tablecomments{Candidate EXors are boldfaced..}
\tablenotetext{a}{~~Coordinates are from the C2D {\it Spitzer} catalog.}
\tablenotetext{b}{~~Separation between the C2D coordinates and those of the identified counterpart.}
\tablenotetext{c}{~~Region as indicated in the C2D catalog.}
\tablenotetext{d}{~~Distance references: (1) for PER, LUPIII, OPH, SER, \#17; (4) for \#15; (6) for \#16 and 18; (7) for \#34; (8) for \#14}
\tablenotetext{e}{~~Identification given in the SIMBAD database.}
\tablenotetext{f}{~~Lum$_{IR}$ is calculated between 1 and 24 $\mu$m.}
\tablenotetext{g}{~~Ratio of IR luminosity (computed between 1 and 3 $\mu$m) to luminosity of a median K5-M5 photosphere (over the same wavelength range).}
\tablenotetext{h}{~~Core \#98 in (4).}

\end{deluxetable}

\begin{deluxetable}{cccccccccccccc} \tablecaption{Photometry the selected EXor candidates. \label{phot:tab}} 
\tabletypesize{\scriptsize} 
\tablewidth{0pt}
\tablehead{ 
\#    &      J   &   	 H   &    K	       &	   W1	   &	IR1	   &	  IR2	 &   W2     &	   IR3  & 	 IR4	   &	 W3	    &   W4	  &	   MP1	&     MP2	\\
      &1.23$\mu$m&1.66$\mu$m &2.16$\mu$m   & 3.4$\mu$m     & 3.6$\mu$m &  4.5$\mu$m  &4.6$\mu$m & 5.8$\mu$m &    8.0$\mu$m & 12$\mu$m   & 22$\mu$m& 24$\mu$m& 70$\mu$m
}
\startdata	 
 12   & $>$17.6  &   15.71   &    13.61	   &	 12.46     &  10.81	   &      9.54	 &  10.65   &	   8.52 &	 7.40      &	 6.62	&   3.74  &   3.22 	&     0.14  \\ 		
 31   &	15.43    &   13.85   &    12.73	   &	 12.04	   &  11.23    &      10.64	 &  10.92   &	  10.22	&    9.33	   & 	 8.70	&   5.48  &   5.16	&     2.55  \\		
 140  & \nodata  &   \nodata &    \nodata  &	 13.96	   &  12.66    &      11.07	 &  11.55   &	  10.40 &    9.28      & 	 7.89 	&   2.91  &   1.65	&    -2.10  \\		
 191  &	16.31    &   14.71   &    13.70	   &	 12.51	   &  12.57    &      11.83	 &  11.59   &	  11.11 &    9.87      &     8.92   &   6.16  &   6.10	&	\nodata \\		
 192  &	12.32    &   10.78   &     9.54	   &	 7.56	   &   8.74    &       7.57	 &   6.66   &	   6.95 &    6.01      & 	 4.66 	&   2.64  &   2.93 	&     0.57  \\		
 354  &	13.73    &   12.14   &    11.08	   &	 9.75	   &   9.84    &       9.19	 &   8.86   &	   8.64 &    7.73      & 	 6.65 	&   4.44  &   4.70 	&	\nodata \\		
 728  &	$>$17.0  &   15.65   &    12.56	   &	 10.99     &  10.18	   &	   9.36	 &   9.70   &	   8.76 &    8.24	   &	 8.12 	&   4.78  &   4.71 	&	\nodata \\	
 828  &	$>$18.2  &   15.79   &    13.87	   &	 12.57     &  11.15	   &      10.06	 &  10.59   &	   9.13 &    8.06	   &	 7.46 	&   4.31  &   4.12 	&    1.295  \\		
 999  &	17.00    &   15.07   &    12.74	   &	 11.23	   &  10.48    &       9.45	 &   9.68   &	   8.64 &    7.77      & 	 6.64 	&   3.56  &   3.60 	&	\nodata \\
 1009 & $>$18.2  &   16.25   &    14.65	   &	 12.98     &  12.27	   &      11.36	 &  11.52   &	  10.74 &    9.84      &	 8.84 	&   5.22  &   5.10 	&    1.94   \\		
 1247 &  \nodata &   \nodata &    \nodata  &     13.91	   &  11.74    &      10.58	 &  11.38   &	   9.67 &    8.87      & 	 8.53 	&   5.47  &   5.27 	&    1.79   \\		
\enddata
\end{deluxetable}
\begin{deluxetable}{ccccccccccc} \tablecaption{Physical properties of the selected EXor candidates. \label{epsilon:tab}} 
\tabletypesize{\scriptsize} 
\tablewidth{0pt}
\tablehead{
\#    &     Incl.      &  A$_{V(IS)}$ & A$_{V(CS)}$                & M$_{star}$      & T$_{star}$    & M$_{env}$                     & L$_{bol}$  \\
      &     (deg)      & (mag)        & (mag)                      &(M$_{\sun}$)     & (K)           &(M$_{\sun}$)                   &(L$_{\sun}$) 
}
\startdata
 12        & 81-18     & 32-38       & 8-4.3~10$^{-4}$             & 0.4-1.6         & 3518-4859     &2.4~10$^{-4}$-1.2~10$^{-8}$    & 1.9-3.0 \\
 31        & 81-75     & 5.9-4.2     & 106-29                      & 0.6-0.2         & 3844-3021     &7.0~10$^{-8}$-8.8~10$^{-6}$    & 0.8-0.3 \\
 140       & 63-63     & 26-49       & 733-271                     & 0.1-0.2         & 2755-3064     &7.3~10$^{-2}$-1.4~10$^{-2}$    &1.1-2.3 \\
 191       & 81-81     & 3.4-2.1     & 105-114                     & 0.2-0.2         & 3050-3062     &5.9~10$^{-9}$-3.9~10$^{-9}$    & 0.2-0.2 \\
 192       & 81-31     & 5-9         & 147-2.8~10$^{-4}$           & 2.1-1.5         & 5718-5209     &1.7~10$^{-6}$-2.0~10$^{-8}$    & 13-5.4   \\
 354       & 41-50     & 9.1-9.1     & 3.6~10$^{-4}$-3.6~10$^{-4}$ & 0.1-0.1         & 2936-2936     &7.3~10$^{-7}$-7.3~10$^{-7}$    & 0.2-0.2 \\
 728       & 63-41     & 40-44       & 1.2-1.6~10$^{-2}$           & 0.3-0.8         & 3433-4100     &5.3~10$^{-3}$-2.8~10$^{-4}$    & 0.9-1.8 \\
 828       & 87-18     & 19-38       & 141-5.6~10$^{-5}$           & 0.3-0.2         & 3491-3085     &3.3~10$^{-4}$-1.7~10$^{-9}$    & 1.0-0.2 \\
 999       & 69-32     & 16-22       & 17-0.8                      & 0.2-0.1         & 3033-2903     &1.4~10$^{-2}$-0.17             & 0.8-0.7 \\
 1009      & 87-32     & 10-41       & $>$1000-1.3~10$^{-2}$       & 1.3-0.3         & 4487-3255     &1.3~10$^{-7}$-1.8~10$^{-3}$    & 1.8-0.5 \\
 1247      & 87-56     & 48-48       & $>$1000-4.6~10$^{-3}$       & 2.3-0.3         & 9019-3472     &1.1~10$^{-8}$-6.5~10$^{-3}$     & 1.8-0.4 \\

\enddata
\tablecomments{In columns 2-8 the first value refers to the best fit, the second to the model with a $\chi^2$ that is 20\% higher than the minimum $\chi^2$.}
\end{deluxetable}

\end{document}